\documentstyle[12pt,epsf]{article}
\newcommand{\beq}{\begin{equation}}
\newcommand{\eeq}{\end{equation}}
\newcommand{\beqa}{\begin{eqnarray}}
\newcommand{\eeqa}{\end{eqnarray}}
\def\nue{{\nu_e}}
\def\anue{{\bar\nu_e}}
\def\numu{{\nu_{\mu}}}
\def\anumu{{\bar\nu_{\mu}}}
\def\nutau{{\nu_{\tau}}}

\newcommand{\dm}{\mbox{$\Delta{m}^{2}$~}}
\newcommand{\st}{\mbox{$\sin^{2}2\theta_V$~}}
\begin{document}

\begin{center}
{\bf Supernova constraints on neutrino mass and mixing}

\vskip 10pt

Srubabati Goswami,
Physical Research Laboratory, Ahmedabad 380009, INDIA

\vskip 10pt
{\bf Absrtact}
\end{center}
\vskip 10pt
In this article I review the constraints on neutrino mass
and mixing coming from type-II supernovae.  The bounds obtained on
these parameters from shock reheating, r-process nucleosynthesis and
from SN1987A are discussed.  Given the current constraints on neutrino
mass and mixing the effect of oscillations of neutrinos from a nearby
supernova explosion in future detectors will also be discussed.

\section{Overview of type-II supernova}
Massive stars with $8M_{\odot} \leq M \leq 60M_{\odot}$ shine for
$10^7$ years via thermonuclear burning producing successively
$H,He,C,O$ and so on. The end product of this chain is $^{56}{Fe}$.
Since $Fe$ has the highest binding energy per nucleon the
thermonuclear reactions stop at the center and the the pressure ceases
to have a part coming from radiation.  At this stage the star has a
``onion-skin" structure.  As the mass of the core becomes greater than
the Chandrasekhar limiting mass the pressure of the relativistic
electron gas alone can no longer counterbalance the inward
gravitational pressure.  The collapse is triggered off by the
photodissociation of $Fe$-nuclei and/or electron capture which reduces
the electron gas pressure further. As the collapse proceeds the core
density rises, causing an increase in the electron chemical
potential. Subsequently the electron Fermi energy becomes higher than
the capture threshold and facilitates electron capture by nuclei and
free protons leading to the neutronization of the core. This further
reduces the e$^{-}$ pressure thereby accelerating the collapse. This
is known as the infall stage \cite{bethe}.  When the core density
becomes of the order of supranuclear densities ($10^{14} {\rm g}/cc$) the
infall is halted and the infalling material bounces back.  The outer
core still continues its infall.  The collision of this rebounding
inner core with the infalling outer core results in the propagation of
a shock wave into the mantle. This shock wave is believed to be
instrumental in causing the supernova explosion. The inner core
develops into the `proto-neutron star'.

\subsection{Neutrino Production and Trapping}
There are three distinct phases of 
neutrino emission from a type-II  supernova. 
\begin{itemize}
\item the infall phase
\item the prompt $\nu_e$ burst
\item the thermal emission phase
\end{itemize}
During the infall stage, mainly electron neutrinos  
are produced via `neutronization' reactions:
\begin{center}
(i) e$^{-}$ + (Z,A) $\rightarrow$ $\nu$ + (Z-1,A)\\
(ii) e$^{-}$ + p $\rightarrow$   $\nu$ + n
\end{center}
As positrons are much fewer, the corresponding antineutrinos cannot
be produced by similar reactions. Also, since the $\mu$ and $\tau$
leptons present are negligible in number, charged current interactions
leading to the production of $\nu_{\mu}$ and $\nu_{\tau}$ can be
neglected. The thermal processes that yield $\nu\overline{\nu}$ pairs
of all flavors are largely suppressed while the infall proceeds,
since the temperature is not high enough.

Initially these neutrinos escape freely from the
star but subsequently the weak interaction of these neutrinos with
nuclei and  nucleons inhibits such free-streaming and neutrinos
transport outwards by diffusion.
The transport of neutrinos outwards has been considered using different
detailed schemes \cite{bruen}. A semianalytic approach adopted in 
\cite{BBAL}. 
uses neutrino
diffusion equation,
\begin{equation}
{\frac{\partial{ n}_{\nu}}{\partial{ t}}} \; =
\; {\frac{1}{{ r}^{2}}} \: {\frac{\partial}{\partial { r}}} \:
\left( \; { r}^{2} \; {\frac{1}{3}} \; { c} \; \lambda_{\nu}
\; {\frac{\partial { n}_{\nu}}{\partial { r}}} \; \right)
\label{diff}
\end{equation}
where $\lambda_\nu$ is the neutrino mean free path given by \cite{lamb}
\begin{equation}
\lambda_\nu~=~1.0 \times 10^{6}{{\rho}_{12}}^{-1}~\left(
\frac{1}{12} X_{h}{\bar{A}} ~+~
X_{n}\right) ^{-1}~\left( \frac{E_{\nu}}{10~\rm  MeV}\right) ^{-2} ~\rm {cm}.
\label{lwv}
\end{equation}
X$_{h}$ and X$_{n}$ are the fractions by mass of heavy nuclei and
neutrons. $\rho_{12}$ is the density of stellar material in 10$^{12}$
{\rm g}/cc. 
Using equation (\ref{diff}) it was shown that the neutrinos diffuse out
of the 
material in about $\frac{1}{9}$ sec. This is much larger than the
hydrodynamic time scale of collapse which is of the order of
1 millisecond. This indicates that the $\nu_{L}$'s are effectively
trapped within the core during the collapse.

Subsequently they are emitted from a ``neutrino-sphere" which is defined
as the radius from where the neutrinos can escape freely:
\begin{equation}
\int^{\infty}_{R_\nu}\frac{dr}{\lambda_\nu} = \frac{2}{3}
\end{equation}
where $\lambda_\nu$ is given by eq. (\ref{lwv}). 

The prompt $\nu_e$ burst
is emitted when the shock wave (being formed  at a 
distance $\sim$ 20 km) passes through the neutrino-sphere (which is
at a distance of $\sim$ 50 km at this epoch). 
The passage of the shock dissociates the $^{56}{Fe}$ nuclei: 
\begin{center}
$^{56}Fe \rightleftharpoons 13\alpha + 4n$\\
$\alpha \rightleftharpoons 2p +2n$.
\end{center}
The neutrinos are produced via e capture on these protons causing this
prompt $\nu_e$ burst. However most of the neutrinos still remain trapped
within the inner core. 

In the shock heated regions, thermal processes produce
$\nu\overline{\nu}$ pairs. 
\begin{center}
$e^{+} e^{-} \rightarrow \nu + \bar\nu$ (pair production)\\
$N~ + N~ \rightarrow~ N~ + N~ + \nu~ + \bar{\nu}$~ (nucleon-nucleon
brehmstraullang)
\end{center}
Neutrinos and antineutrinos of all flavours are produced by these
processes. 

\section{How does supernova explode?}
For stars in the mass range $8M_{\odot} \leq M \leq 15M_{\odot}$,
under some very special conditions on the size and structure of the
core and the equation of state,
the shock continues its outward propagation and the star explodes within
some tens of milliseconds after the beginning of collapse.
This is the prompt explosion scenario \cite{prompt}.
For more massive stars the energy of the shock gets dissipated in
dissociating nuclei and producing $\nu \overline{\nu}$ pairs and the shock
stalls at a radius of a  few hundred kms and becomes an accretion
shock. It is subsequently revived by the heating caused by neutrinos
from the neutrino-sphere. This is the delayed explosion mechanism
\cite{wilson,bw85}. 
However this mechanism generates a feeble shock with energies 
less than $10^{51}$ ergs, a factor of 3 to 4 less than observed values.

Various mechanisms are considered which can generate a successful shock.
It is now realised that 
convection of matter in the core plays a crucial role in the energy
transport \cite{bethe95} and hydrodynamic calculations in two or three
dimensions are being pursued by different groups \cite{janka}.
Other important mechanisms like
improved -supernova conditions, soft equation of state, general
relativity at high densities,  and improved neutrino
physics have all been invoked to solve this problem. 
Fuller {\em{et al.}} \cite{full}
pointed out that  matter-enhanced resonant flavor conversions of
neutrinos in the region between the neutrino-sphere and the stalled
shock can increase the shock heating rate appreciably and can result in
a delayed explosion with energy  $\geq 10^{51}$ ergs, for a
cosmologically significant $\nu_{\mu}$ or $\nu_{\tau}$ mass of 10
-- 100 eV and small vacuum mixing angle. 
The basic idea 
is that due to neutrino flavor mixing $\nu_{\mu}$s or $\nu_{\tau}$s
get converted to $\nu_e$s. 
Since $\nu_\mu$ and $\nu_\tau$ undergoes only neutral current
interactions whereas $\nu_e$ undergoes both neutral current and charged
current interaction 
the average energy of
$\nu_{\mu}$ or $\nu_{\tau}$ is higher than that of $\nu_e$. Thus the
upshot of such flavor conversion is production of higher
energy $\nu_e$s which can heat the shock more effectively.
Since $\nu_e$'s are assumed to undergo MSW resonance the
$\bar\nu_e$s remain unaffected.

The late time neutrino  heating of the shock
is caused by their absorption reactions on the nuclei as well as on free
nucleons  and by charged and neutral current scattering reactions. 
In this scenario the energy absorbed by matter
behind the shock front/{\rm g}/sec assuming the matter to be nucleonic 
is given by \cite{bw85},
\begin{eqnarray}
\dot{E} = (223 {\rm MeV}/ {\rm{nucleon~sec}})
               \frac{1}{{R_7}^2}\left[X_p L_{52}
                 ({\overline\nu_e})
                 \left({\frac{T_{\overline\nu_e}}{5 {\rm MeV}}}\right)^2
              + {X_n} L_{52}(\nu_e)
                 \left({\frac{T_{\nu_e}}{5 {\rm MeV}}}\right)^2\right]
\end{eqnarray}
where $R_{7}$ is the radial distance from the center of the star
units of $10^7$ cm, $L_{52}$ denotes the neutrino luminosity in units
of $10^{52}$ ergs/sec, $X_{n}$ and $X_{p}$ are the neutron and proton
fractions respectively and $T_{\nu}$s are the temperatures of the
respective neutrino-spheres.
In presence of complete flavour conversion between $\nu_e$ and $\nu_\tau$ 
the $T_{\nu_e}$ in the above equation would be changed to $T_{\nu_\tau}$.
One can make a rough estimate of the increase in heating rate assuming 
the luminosities to be equal when one obtains
\begin{equation}
\frac{\dot E_{\rm osc} } { \dot{E} } = X_{p} + X_{n}
\left( \frac{T_{\nu_\tau}}{T_{\bar\nu_e}} \right)^2 \,.
\label{eoscbye}
\end{equation}
Taking some typical values,
$T_{\nu_\tau} \approx $ 7 MeV, $T_{\bar\nu_e} \approx $ 5 MeV,
$X_{n} \approx $ 2/3 and $X_{p}\approx1/3$ one obtains
$\dot E_{\rm osc}/\dot{E}=1.64$.  
However this is only approximate. 
In practice one has to also account for the rate of energy loss due to
radiation and incorporating all these a detail numerical simulation shows a
60\% increase in the explosion energy \cite{full}.

\subsection{Parameters for complete MSW flavour conversion of neutrinos}
Neutrino flavor evolution in supernovae has
been discussed in detail in \cite{full,sigl,full95}. Here we give the
basic equations needed for our purpose. 
We confine our discussions to two neutrino flavors and take these to
be  $\nu_e$ and $\nu_{\tau}$. 
The mass matrix ${M_{F}}^{2}$ (see eq.(3.4)) in flavor basis
for this case is  
\begin{equation}
{M_{F}}^{2} = U {\pmatrix {0 & 0   \cr  0 & \Delta  \cr }} U^{\dagger}
+ {\pmatrix{ V_{cc} & 0 \cr 0 & 0 \cr}} + {\pmatrix{ V_{nc} & 0 \cr 0 &
V_{nc} \cr}} + {\pmatrix{ V_{\nu_e \nu_e} & V_{\nu_e \nu_\tau} \cr V_{\nu_\tau
\nu_e} & V_{\nu_\tau \nu_\tau} \cr}}
\label{mf}
\end{equation} 
$V_{cc}$ is due to charged current $\nu_e-e$ scattering and is given
by $\sqrt{2}G_{F}n_{e}E$, $n_e$ is the net electron density. $V_{nc}$ is
due to neutral current scattering off neutrons and nuclei. This is
identical for both flavors and can be discarded. The last piece in
(\ref{mf}) is due to neutrino-neutrino exchange scattering. 
It has been shown in \cite{full95} that this term has negligible effect
on the adiabatic transitions important for  the shock-reheating epoch
and in what follows we will neglect this contribution. 
By equating the two diagonal
elements of the remaining terms in (\ref{mf}) one arrives at the
following resonance condition \cite{full}, 
\begin{equation}
\rho_{\rm res} = (6.616 \times 10^{6} {\rm g}/cc)
\left(\frac{\cos{2\theta_{V}}}{Y_e}\right) 
\left(\frac{\Delta{m^2}}{{\rm eV}^{2}}\right) 
\left(\frac{1 {\rm MeV}}{E_{\nu}}\right) 
\label{resden}
\end{equation}
The effectiveness of neutrino oscillations in increasing the heating
rate depends very crucially on whether a resonance is encountered
between the  neutrino-sphere and the shock front or not.  
If a neutrino of flavor $f$ encounters a resonance between the
neutrino-sphere and the shock then its probability to remain a $\nu_{f}$
at the position of the shock is given as,
\begin{equation}
P(\nu_f \rightarrow \nu_f) = 0.5 + (0.5 -
P_J)\cos{2\theta_{R_\nu}}\cos{2\theta_{R_{m}}}
\label{pnuf}
\end{equation}
where $\theta_{R_\nu}$ ($\theta_{R_{m}}$) is the neutrino mixing angle 
in matter at the position of the neutrino-sphere (shock) and can be
expressed as,
\begin{equation}
\tan{2\theta_{R_{i}}}~=~{\tan2\theta_{V}}/{(1- \rho_{R_{i}}/\rho_{\rm res})}
\label{tan2ri}
\end{equation}
where $R_{i}$ can be $R_{\nu}$ or $R_{m}$.
$E_{\nu}$ is the neutrino energy; 
$P_{J}$ is the non-adiabatic transition probability between the
two neutrino states and can be expressed in the
Landau-Zener approximation as,
\begin{equation}
P_{J} = exp(-{E_{NA}}/{E_{\nu}})
\label{pj}
\end{equation}
where, 
\begin{equation}
E_{NA} = \frac{\pi}{2}
\left(\frac{{\sin^2}2{\theta_{V}}}{\cos2\theta_{V}}\right) 
\frac{\dm}{({\frac{1}{n_e}}\frac{d{n_e}}{dr})_{\rm res}}
\label{ena1}
\end{equation} 
A neutrino passing through the resonance density can
undergo complete flavor transformation for appropriate values of the
parameters. A total conversion of the $\nu_{\tau}$, which carry a
higher energy, to $\nu_{e}$
between the neutrino-sphere and the shock will generate maximum
heating.
In the region between the neutrino-sphere and the shock the neutrinos 
pass through a decreasing density profile. 
For fixed values of ${\Delta{m}}^{2}/E$ and $\theta_{V}$, (\ref{resden})
allows one to determine the resonance density. 
Since we are interested in complete conversion we have to ensure that
not only is a resonance attained, but $P(\nu_f \rightarrow \nu_f)$ as
given by (\ref{pnuf}) is zero. 
Following situations might arise:
\begin{itemize}
\item $\rho_{\rm res} > \rho_{R_{\nu}} > \rho_{R_{m}}$, which implies that 
resonance position is below the neutrino-sphere. 
No resonance is
achieved by the neutrinos coming from the
neutrino-sphere and hence the
probability of level-jumping at resonance $P_{J}$ is zero in the region
between the neutrino-sphere and the shock. Considering the limiting case of 
$\rho_{\rm res} \gg  \rho_{R_{\nu}}$, 
(\ref{tan2ri}) implies $\theta_{R_{\nu}}
\rightarrow \theta_{V}$. 
Also, $\rho_{\nu} \ll   \rho_{R_m}$ and $\theta_{R_m} \rightarrow \theta_V$.
Then from (\ref{pnuf}), $P_{\nu_f 
\nu_f} \rightarrow \frac{1}{2}(1+ \cos^2{2\theta_{V}})$. The maximum
conversion that one can obtain is therefore 0.5 when $\cos 2\theta_{V}
\approx$ 0.

\item $\rho_{\rm res} < \rho_{R_{m}} < \rho_{R_{\nu}}$, which implies a
resonance position outside the shock and again $P_{J}$ would be 0. 
In the limit $\rho_{\rm res} \ll  \rho_{R_{m}}$,
This case
corresponds to $\theta_{R_m} \rightarrow \pi/2$. 
$\theta_{R_{\nu}} 
\rightarrow \pi/2$ as well so that from (\ref{pnuf}),
$P_{\nu_f \nu_f} \rightarrow 1$. Thus in this situation complete
conversion is not a possibility.

\item $\rho_{R_{\nu}} \geq \rho_{\rm res} \geq \rho_{R_{m}}$, which implies
a resonance is met between the neutrino-sphere and the shock. 
The limiting case $\rho_{\rm res} \ll  \rho_{R_{\nu}}$ implies
$\theta_{R_{\nu}} \rightarrow \pi/2$.
On the other hand in the limit $\rho_{\rm res} \gg  \rho_{R_{m}}$ corresponds
$\theta_{R_m} \rightarrow \theta_V$ so that
(\ref{pnuf}) becomes $0.5 - (0.5-P_J)\cos{2\theta_{V}}$.
For small $\theta_{V}$, $\cos2\theta_{V} \rightarrow 1$ and 
$P_{\nu_f \nu_f} \approx P_{J}$. Complete conversion can be obtained if 
$P_{J}$ is zero {\it i.e.} transitions are adiabatic. If on the
other hand $\theta_{V}$ is large and $\cos2\theta_{V}$ is 0,
$P_{\nu_f \nu_f} \approx$ 0.5 and complete conversion is not possible. 

\end{itemize}

Thus complete conversion is possible only in the last situation
discussed above which requires $\rho_{R_{\nu}} \geq \rho_{\rm res} \geq
\rho_{R_{m}}$.  
This gives
the following constraint on ${\Delta{m}}^2$ for 
$\cos{2\theta_{V}} \approx $ 1,
\begin{eqnarray}
(5.6 \times 10^{-8}\, {\rm eV}^2) \left(\frac{E_{\nu}}{{\rm MeV}}\right)
\left(\frac{ \rho_{R_{m}} }{{\rm g/cm}^3}\right)   \leq \dm
\leq  (5.6 \times 10^{-8}\, {\rm eV}^2) \left(\frac{E_{\nu}}{{\rm MeV}}\right)
\left(\frac{ \rho_{R_{\nu}} }{{\rm g/cm}^3}\right)
\label{massbound}
\end{eqnarray}
where we have taken $Y_{e} = 0.37$. 
For typical values of the densities and energies this gives 
$\Delta m^2$ $\sim$ 100 - $10^{4} {\rm eV}^2$ which is in the 
cosmologically interesting range.

Complete flavor conversion occurs when transitions are adiabatic. 
From the validity of the adiabatic condition, $E_{NA} \geq E_{\nu}$ 
a lower bound on $\theta_{V}$ can be obtained 
\begin{equation}
\frac{{\sin^2}2{\theta_{V}}}{\cos2\theta_{V}} \geq 1.258 \times 10^{-4}
\left(\frac{{\rm eV}^2}{\dm}\right)~
\left(\frac{{({\frac{1}{n_e}}\frac{d{n_e}}{dr})_{\rm res}}}{{\rm
km}^{-1}}\right)~ 
\left(\frac{E_{\nu}}{{\rm MeV}}\right)
\label{lb}
\end{equation}

Taking some typical values $\Delta m^2 \sim$ 1600 ${\rm eV}^2$ and 
the density scale height $d ln n_e/dr$ $\sim$ 50 km 
one gets
\begin{equation}
\sin^2 2\theta_V \geq 10^{-8} \frac{E_\nu}{10 {\rm MeV}} \,.
\end{equation}
In fig. 1 we give an illustrative plot
taking a density profile \cite{bethe}\\
$\rho=(10^{31}{\rm {\rm g}/cc})(r/{\rm 1 cm})^{-3}$.  
It is to be noted that the actual density profile after collapse does
not follow such a simple power law behaviour. 
The allowed regions in the ${\Delta{m}}^2 - \sin^2 2\theta_V/\cos 2 \theta_V$ plane, 
consistent with (99 -- 100)\% flavor conversion for a typical neutrino
energy of 20 MeV, is shown in fig. 1. This curve is for a
25$M_{\odot}$ star 
with
the shock positioned at the minimum distance from the neutrino-sphere
190 km 
since at this position most stringent constraints on the
parameters are obtained \cite{ska}.

\section{r-process nucleosynthesis}
Heavy neutron rich nuclei beyond the iron group are synthesised by neutron
capture. There are two basic processes:
\begin{itemize}
\item{\bf s-process} or slow process for which the time of neutron capture
$t(n,\gamma) \gg  t_\beta$, where $t_\beta$ is the beta decay life time. 
Thus nuclides are built along the stability valley. 
\item{\bf r-process} or rapid process for which $t(n,\gamma) \ll  t_\beta$
and very neutron rich unstable nuclei are built. 
The above condition requires a neutron density $n_n > 10^{19} cm^{-3}$. 
\end{itemize}
Many authors conjectured that type-II supernovae can be a possible
site for r-process nucleosynthesis 
since it  has the required
high neutron number densities $> 10^{20} cm^{-3}$,    
temperatures $\sim 2-3\times{10^9}^0 K $ and time scales 
$\sim 1$\,s \cite{five}. 
But where exactly in the supernova does the r-process actually take place is 
a debatable issue. In  
the recent years the neutrino heated ejecta from the post core bounce 
environment of a type II supernova or the ``hot bubble"  has been
suggested as a site for r-process \cite{six,ww}.  
The ``hot bubble" is the region between a protoneutron star and the
escaping shock wave in a core-collapse supernova. 
The material in this region has a low density because of the successful
explosion and yet very hot $\sim$ 10$9^{0}$K.
The shock reheating epoch is between $\sim$ 0.1 - 0.6 s after core
bounce whereas the r-process epoch is $\sim$ 3 - 20 s after core
bounce. 
The major advantage which the ``hot bubble" has over other proposed sites 
is that it correctly predicts that only $10^{-4} M_\odot$ of r-process 
nuclei are ejected per supernova \cite{mc}.
The late time ($t_{pb}$= 3 -15s) evolution of $20 M_\odot$ delayed SN
explosion model gives an excellent fit to the solar r-process abundance
distribution \cite{ww}.

For r-process to take place in supernova
neutron rich conditions are needed. This in turn requires  
that the electron fraction $Y_e$ defined as 
\begin{equation} 
Y_e = \frac{ No.~ of~ electrons}{No.~ of~ Baryons} = Y_{p}
\end{equation}
be $< 0.5$ at the weak
freeze-out radius ($r_{\small {WFO}} \sim$ 40 - 100 km). 
$r_{\small {WFO}}$ is defined as 
 radius where the absorption of $\nue$ and $\anue$ on 
free nucleons ($\nu_e n \rightarrow e^{-} p~, \bar\nu_e p \rightarrow
e^{+} n$) freeze out  and is found to be very close to the nuclear 
freeze-out radius in most supernova models.
The expression for the value of $Y_e$ at freeze out is given by Qian et
al. \cite{qian} 
 as 
\begin{equation}
Y_e \approx 
{{1}\over{1+\lambda_{\anue p}/\lambda_{\nue n}}}
\label{ye}
\end{equation}
Where $\lambda_{\nue n}$ and $\lambda_{\anue p}$ are the reaction rates. 
The reaction rate $\lambda_{\nu N}$, where N can be either p or n 
is given by 
\begin{equation}
\lambda_{\nu N} \approx {{L_{\nu}}\over{4\pi r^2}} 
{{\int_0^{\infty} \sigma_{\nu N} (E) f_\nu (E) dE}\over{\int_0
^{\infty} Ef_\nu(E) dE}}  
\label{lamb}
\end{equation}
where $L_\nu$ is the neutrino luminosity (we consider identical luminosity 
for all the neutrino species), $\sigma_{\nu N}$ is the reaction cross-section  
and $f_\nu (E)$ is the normalised Fermi-Dirac 
distribution function with zero chemical potential 
\begin{equation}
f_\nu(E)={{1}\over{1.803 {T_\nu}^3}} {{E^2}\over{exp(E/T_\nu)+1}}
\end{equation}
where $T_\nu$ is the temperature of the particular neutrino concerned. The 
cross section is approximately given by \cite {full} 
\begin{equation}
\sigma_{\nu N} \approx 9.23\times 10^{-44}(E/{\rm MeV})^2 cm^2
\end{equation}
If we calculate $\lambda_{\nue n}$ and $\lambda_{\anue p}$ using eq 
(\ref{lamb}) then 
the expression for $Y_e$ becomes
\begin{equation}
Y_e \approx {{1}\over{1+T_{\anue}/T_{\nue}}}
\end{equation}
Typical values for the neutrino temperatures when r-process is operative are 
\cite {qian}, $T_{\nue}$ = 3.49 MeV, $T_{\anue}$ = 5.08 MeV and $T_{\numu}$ 
=7.94 MeV so that $Y_e \approx 0.41$. 
This being less than 0.5 neutron rich conditions persist 
in the hot bubble and r-process is possible. 

\subsection{MSW transitions and r-process}

Qian {\it et al.}  made a two flavor  
analysis of the matter-enhanced level crossing between $\nu_e $ and
$\nu_\tau $ 
or $\nu_\mu $. They considered a mass spectrum in which $m_{\nu_{\tau,\mu}} 
>m_{\nu_e}$ so that there is resonance between the neutrinos 
only and not between the antineutrinos. 
As a result the more energetic $\nu_{\mu,\tau}$
 ($<E_{\nu_\mu}>$ = $<E_{\nu_\tau}>$ $\sim 25 {\rm MeV})$ get converted to $\nu_e $ 
($<E_{\nu_e}>\sim 11 {\rm MeV}$) increasing the average energy of the
electron neutrinos.  
As a result of this flavour conversion
the neutrino energy distribution function 
itself will change to (assuming two flavors)
\begin{equation}
f_{\nue}^{osc}(E) = P_{\nue\nue}f_{\nue}(E) + P_{\numu\nue}f_{\numu}(E)\\
\label{f21}
\end{equation}
But the antineutrinos do not undergo any transition in this picture 
so that their energy ( $<E_{\bar\nu_e}>\sim
16 {\rm MeV}$) and distribution function remains the same 
\begin{equation}
f_{\anue}^{osc}(E) = f_{\anue}(E) 
\end{equation}
The resonance condition is as given by eq.(\ref{resden}) and 
the mass of the $\nu_{\tau}$ (or $\nu_\mu)$ required to undergo MSW
resonance between the neutrino-sphere and the weak freeze-out radius
was shown to be between 2 and 100 eV which
is the right range for neutrinos to be the hot dark matter of the 
Universe. The densities at the neutrino-sphere and the weak-freeze out radius 
is such that the matter modified mixing angle at the neutrino-sphere tends to
$\pi/2$ and that at the $r_{\small{WFO}}$ $\sim$ the vacuum mixing angle
$\theta_V$ so that the 
The transition probability for small vacuum mixing angles can be 
approximated as 
\begin{equation}
P_{\nu_\tau \nu_e} \approx 1 - P_{J}
\end{equation}
with $P_J$ given by eqs. (\ref{pj}) and (\ref{ena1}).
The typical value of the density scale height is now $\sim$ 0.5 km
which is two orders of magnitude smaller than the value in the shock
reheating case. So that the lower limit on $\sin^2 2\theta_V$ is about
two orders of magnitude larger than it was for the shock revival
scenario. 
The detail analysis of \cite{qian} gives the curve in fig.2. 
The area to the left of the curve is consistent 
with the $Y_e < 0.5$ constraint required for r-process. 

\section{Neutrinos from SN1987A}
From the supernova SN1987A 11 neutrino induced events were 
detected by the Kamiokande
detector and 8 events by the IMB detector \cite{k2,imb}. These
detectors are water Cerencov detectors. 
The dominant reaction for the detection of SN neutrinos are the charged
current $\bar\nu_e p \rightarrow n e^{+}$. 

\subsection{$\anue$ oscillation}
Neutrino oscillation can modify the signal at the detector 
following manner  
\begin{equation}
F(\bar\nu_e) = F_{0}(\nu_e)~(1-p)~+~p F_{0}(\bar\nu_\mu)
\label{spectrum}
\end{equation}
where $F_{0}$ is the original spectrum and $F$ denotes the observed spectrum. 
If p =1 then the detected spectrum is same as the original spectrum and if 
$p<1$ the detected $\bar\nu_e$ spectrum is a mixture of the original 
$\bar\nu_e$ and $\bar\nu_\mu$ spectrum. 
Comparing the observed energy spectra and the expected spectra in \cite{sb} 
it was deduced that $p\leq 0.35$ (99\% C.L.) depending on the assumed
primary neutrino spectra.  
The above constraint on $p$ can now be used to constrain the oscillation
parameters. 

In \cite{sb} a normal mass hierarchy ($m_\numu > m_\nue$)
between the neutrino states was considered. For this case resonance occurs
between the neutrinos and not the antineutrinos. 
There can still be some transitions between the antineutrino states 
if the vacuum mixing angle is large. How much conversion is consistent
with the observation is determined by the constraint on the  
permutation factor $p$ and  they obtained the following upper bound on
the mixing angle 

\[ \sin^2 2\theta_V  \leq  \left\{ 
\begin{array}{ll}
0.9 & \dm \gg  10^{-9} {\rm eV}^2 \\
0.7 & \dm \ll  10^{-9} {\rm eV}^2
\end{array}
\right. \]

Thus the vacuum oscillation solution  and
partly the large mixing MSW solution to the solar neutrino problem is
disfavoured. 
However the above bounds are sensitive to the neutrino spectrum predicted
by the theoretical supernova models. 
In \cite{kk}  certain predictions about the signal characteristics were
disregarded and they arrived at the opposite conclusion that large
mixing angles are actually favoured. But their conclusion is valid only
for \dm $\leq 10^{-10} {\rm eV}^2$. 
In \cite{jnr} a re-examination of the above has been done and they reach
the same conclusion as in \cite{sb} that the solar vacuum solution is 
incompatible with the SN1987A data if the predicted spectrum shape is
assumed to be correct. 

If one takes an inverted mass hierarchy with $m_\nue > m_\numu$ the
$\anue-\anumu$ transitions are resonant and there is a large amount of
conversion inconsistent with observation. 
A large range of mass and mixing parameters 
\begin{center}
$10^{-8} \leq \dm \leq 10^{4}$ \\
$10^{-8} \leq \st \leq 1$
\end{center}
are thus excluded for an inverted mass scheme from SN1987A
observations. 

\subsection{$\nue$ oscillation}
Since the first event in Kamiokande shows forward peaking there was a
speculation that it might come
from $\nue-e$ scattering and many authors discussed the impact of
matter-induced oscillation on the basis of this \cite{ja} and 
a large range of mass and mixing angles can be shown to be disfavoured. 
However because these analyses are based on a single event 
the statistical significance is questionable. 

Haxton in 1987 pointed out that the reaction ${^16}{O}(\nue,e)F^{16}$
cross-section increases very sharply with energy \cite{haxton}. 
Thus this reaction which occurs in the water Cerencov detectors 
can be very sensitive to neutrino flavour conversion
as the $\nue$s coming from flavour converted $\numu$s or $\nutau$s will
have higher energy.
In \cite{qf} the effect of matter enhanced neutrino conversion on 
these events was examined. 
They showed that in KII if the $P_{\nu_e \nu_e}$ is 1 corresponding to no
oscillation the number of oxygen events are 7 whereas for complete 
conversion ($P_{\nu_e \nu_e}$=0) the number of events increase to 32. 
They showed that  a mass spectrum which is consistent with the constraints
of r-process nucleosynthesis in SN and the solar neutrino problem: 
$\Delta_{13} \approx 1 - 10^4 {\rm eV}^2$ and 
$\sin^2 2\theta_{13} = 4 \times 10^{-6}$ and 
$\Delta_{12} \approx 10^{-6} - 10^{-5} {\rm eV}^2$  and $\sin^2 2\theta_{12}$ 
in the non-adiabatic solar range can give a large conversion probability. 
This mass spectrum gives two well separated resonances in the supernova.

\section{Future Detectors}

The next generation solar neutrino detectors like SNO or Super-Kamiokande 
can also be used to detect neutrinos from a nearby supernova explosion.
SK is a 32 kt water cerencov detector. The dominant detection reaction is 
capture of $\anue$ on protons. 
SNO is a heavy water detector made of 1 kton of
pure $\rm D_2O$. 
The main detection reactions are
\begin{center}
$\nu_e + d \rightarrow p+p+e^-\quad({\rm CC\ absorption})$\\
$\nu_x + d \rightarrow p+n+\nu_x\quad({\rm NC\ dissociation})$\\
$\bar\nu_e + d \rightarrow n+n+e^{-} \quad({\rm CC\ absorption}).$ \nonumber
\label{reactionNC}
\end{center}
There have been various attempts to estimate the effect of
non-zero neutrino mass and mixing on the expected neutrino signal
from a galactic supernova \cite{akh,bkg,qf}.
Burrows {\it et al.} \cite{bkg} have considered the effect of
vacuum oscillations for  SNO and have found that
with two-flavours the effect of vacuum oscillations on the signal is
small, using their model predictions for the
different $\nu$ luminosities.
Two recent works have considered  
the effect of two and three generation vacuum oscillation on 
the expected signal in SNO and Super-Kamiokande.
In \cite{majum} the effect of vacuum oscillations on the neutronisation 
$\nu_e$ pulse was considered \cite{majum}. 
This case will be discussed in \cite{majum2}.
In \cite{skd}, the effect of flavour oscillation on the post-bounce thermal
neutrino flux was considered. 
For the mass and mixing parameters they take two
scenarios.  
\begin{itemize}
\item {\bf scenario 1:} Here  threefold maximally mixed 
neutrinos with the mass spectrum $\Delta m_{13}^2 \approx 
\Delta m_{23}^2 \sim 10^{-3} {\rm eV}^2$ corresponding to the atmospheric 
range while $\Delta m_{12}^2 \sim 10^{-11} {\rm eV}^2$ in accordance with 
the solar neutrino problem was considered.  
\item {\bf scenario 2:} Here $\Delta m_{12}^2 \sim 10^{-18} {\rm eV}^2$ 
for which $\lambda \sim L$ and the oscillation effects are observable 
while $\Delta m_{13}^2 \approx \Delta m_{23}^2  \sim 10^{-11} {\rm eV}^2$ 
(solar range).
\end{itemize}
For the latter case the oscillations due to $\Delta m_{13}^2$ 
and $\Delta m_{23}^2$ are averaged out as the neutrinos 
travel to earth.
For $\theta_{13}$ they considered two sets of 
values  
$\sin^2 2\theta_{13}$ = 1.0 (the maximum allowed value) (case 2a) and with 
$\sin^2 2\theta_{13}$ = 0.75, the best fit value from solar neutrino
data (case 2b).
In both these scenarios the matter effects are not important. 
They calculated the number of expected neutrino events 
from a typical type II supernova at a distance of 10kpc
for the main reactions in $\rm H_2O$ and $\rm D_2O$ in presence
and absence of oscillations \cite{skdt} for the various scenarios. 
As a result of mixing the  
$\nu_\mu$  and $\nu_\tau$s   
and the corresponding antineutrinos oscillate 
(with average energy $\sim$ 25 MeV) into 
$\nue$ and $\anue$ during their passage from the  
supernova to the 
detector resulting in higher energy $\nue$ and $\anue$.
Hence all the charged current events show increase in number
compared to the no oscillation scenario. 
For the  $^{16}O$ reaction
the increase is  maximum (by 120 or 130\%). However this is dependent
on the supernova model used \cite{totani}.  

In conclusion, notwithstanding the various uncertainties in theoretical
modeling of supernova it can be used as a good testing ground for new neutrino
properties. In this article the constraints  on neutrino mass
and mixing are discussed from considerations of neutrino oscillation 
in supernova. Constraints can also be obtained on  magnetic moment of
neutrino in case of resonant spin-flavour transitions or on
the strengths of flavour changing neutral current interactions in case
of oscillations of massless neutrinos and other new neutrino properties
from similar considerations.

\begin{figure}[p]
\epsfxsize 15cm
\epsfysize 15cm
\epsfbox[25 151 585 704]{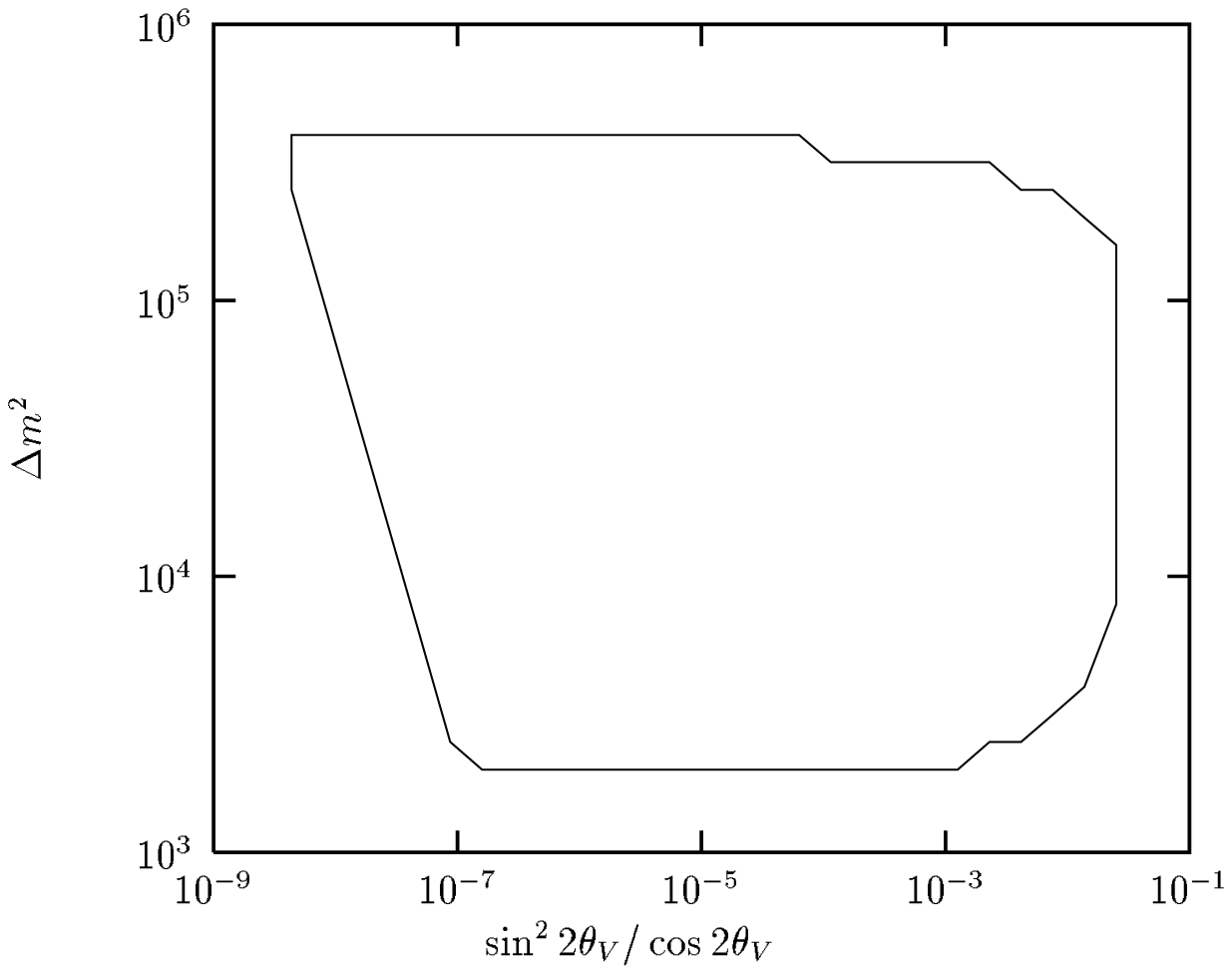}
\vskip -8cm
\caption[Figure~ref{fig1}]{\label{fig1}
The allowed area in the \dm vs. $\sin^2 2\theta_V/\cos2\theta_V$ plane that
gives complete conversion in a $25M_\odot$ star.}
\end{figure}

\begin{figure}[p]
\epsfxsize 15cm
\epsfysize 15cm
\epsfbox[25 151 585 704]{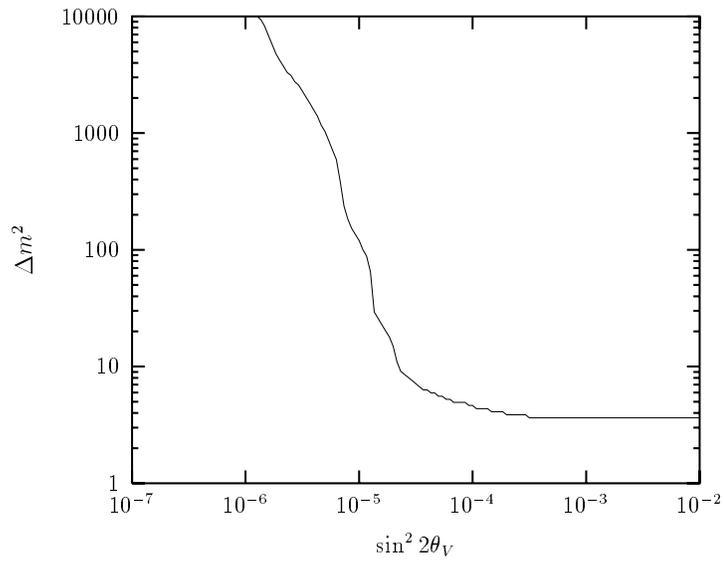}
\vskip -8cm
\caption[Figure~ref{fig2}]{\label{fig2}
The allowed area consistent with the $Y_e < 0.5$ constraint 
is to the left of the solid curve.} 
\end{figure}
\end{document}